\documentclass[universe,article,accept,pdftex,moreauthors]{Definitions/mdpi} 
\firstpage{1} 
\pubvolume{1}
\issuenum{1}
\articlenumber{0}
\pubyear{2022}
\copyrightyear{2022}
\datereceived{18 April 2022} 
\dateaccepted{3 June 2022} 
\datepublished{} 
\hreflink{https://doi.org/} % If needed use \linebreak
\pdfoutput=1

\usepackage{mathtools,amssymb}
\usepackage{aas_macros}
\usepackage{graphicx} %use graph format
\usepackage{epstopdf}
\usepackage{changes}

\Title{Constraints on the Helium Abundance from Fast Radio Bursts}

\TitleCitation{Constraints on the Helium Abundance from Fast Radio Bursts}

\Author{Liang Jing and Jun-Qing Xia *}
\AuthorNames{Liang Jing, and Jun-Qing Xia}

\AuthorCitation{Jing, L.; Xia, J.-Q.}
\address[1]{Department of Astronomy, Beijing Normal University, Beijing 100875, China; 202021160013@mail.bnu.edu.cn}%MDPI: Newly added information, please confirm.

\corres{\hangafter=1 \hangindent=1.05em \hspace{-0.82em} Correspondence: xiajq@bnu.edu.cn}

\abstract{Through the relationship between dispersion measures (DM) and redshifts, fast radio bursts (FRBs) are considered to be very promising cosmological probes. In this paper, we attempted to use the DM-z relationship of FRBs to study the helium abundance ($Y_{\rm He}$) in the universe. First, we used  17 current FRBs with known redshifts for our study. %Please check intended meaning has been retained
	 Due to their low redshifts and the strong degeneracy between $Y_{\rm He}$ and $\Omega_bh^2$, however, this catalog could not provide a good constraint on the helium abundance. Then, we simulated 500 low redshift FRB mock data with $z\in[0,\,1.5]$ to forecast the constraining ability on $Y_{\rm He}$. In order to break the degeneracy between $Y_{\rm He}$ and $\Omega_bh^2$ further, we introduced the shift parameters of the Planck measurement $(R,l_A,\Omega_bh^2)$ as a prior, where $\Omega_bh^2$ represents the baryon density parameter, and $R$ and $l_A$ correspond to the scaled distance to recombination and the angular scale of the sound horizon at recombination, respectively. We obtained the standard deviation for the helium abundance: $\sigma({Y_{\rm He}}) = 0.025$. Finally, we considered 2000 higher redshift FRB data with the redshift distribution of $[0,\,3]$ and found that the constraining power for $Y_{\rm He}$ would be improved by more than \mbox{2 times}, $\sigma({Y_{\rm He}}) = 0.011$, which indicates that the FRB data with high redshift can provide a better constraint on the helium abundance. Hopefully,  large FRB samples with high redshift from the Square Kilometre Array can provide high-precision measurements of the helium abundance in the \mbox{near future}.}


\begin{document}

%%%%%%%%%%%%%%%%%%%%%%%%%%%%%%%%%%%%%%%%%%
%\setcounter{section}{-1} %% Remove this when starting to work on the template.
%\section{How to Use this Template}

%The template details the sections that can be used in a manuscript. Note that the order and names of article sections may differ from the requirements of the journal (e.g., the positioning of the Materials and Methods section). Please check the instructions on the authors' page of the journal to verify the correct order and names. For any questions, please contact the editorial office of the journal or support@mdpi.com. For LaTeX-related questions please contact latex@mdpi.com.%\endnote{This is an endnote.} % To use endnotes, please un-comment \printendnotes below (before References). Only journal Laws uses \footnote.

% The order of the section titles is: Introduction, Materials and Methods, Results, Discussion, Conclusions for these journals: aerospace,algorithms,antibodies,antioxidants,atmosphere,axioms,biomedicines,carbon,crystals,designs,diagnostics,environments,fermentation,fluids,forests,fractalfract,informatics,information,inventions,jfmk,jrfm,lubricants,neonatalscreening,neuroglia,particles,pharmaceutics,polymers,processes,technologies,viruses,vision

\section{Introduction}

Fast radio bursts (FRBs) are very short (ms) transients observed in frequencies from $\sim$100 MHz up to a few GHz \citep{2007Sci...318..777L,2013Sci...341...53T,2015MNRAS.447..246P}.
The triggering mechanisms of FRBs are mysterious and still highly debated, but at least some FRBs can be produced by magnetar engins \citep{2020Natur.587...54C,2019MNRAS.485.4091M,2020MNRAS.494.2385K,2020MNRAS.498.1397L}, also possibly by superconducting strings \citep{2008PhRvL.101n1301V}, or even by mysterious objects concerning strange quark stars \citep{2021Innov...200152G}.
The first FRB was discovered by \citet{2007Sci...318..777L}, and hundreds of FRBs have been observed since then from several radio surveys, such as CHIME \citep{2018ApJ...863...48C}, Parkes \citep{2016PASA...33...45P}, and ASKAP \cite{2007IAUS..240..429J}. In the future, the Square Kilometre Array (SKA) is expected to detect 
$\sim$$10^4$ FRBs per decade \cite{2017ApJ...846L..27F}. Among those discovered FRBs, one repeating burst FRB121102 localized at z $\sim$ 0.19 established the cosmological origin of these events \cite{2016Natur.531..202S}. Furthermore, most FRBs have anomalously large dispersion measures (DM), which are related to their high redshifts. Through the relationship between DM and redshifts, FRBs are considered to be very promising cosmological probes. Several studies have been conducted to forecast the determinations of cosmological parameters using the future mock FRBs, such as the Hubble Constant \cite{2022MNRAS.511..662H,2022MNRAS.tmpL..23W}, the dark energy equation of state \cite{2014ApJ...788..189G}, the fraction of baryon mass in the intergalactic medium (IGM) \cite{2019ApJ...876..146L,Dai:2021czy}, the reconstruction of reionization using FRBs \cite{2021JCAP...05..050D}, and so on. %Precise constraints can be made in the future as the number of probing samples increases.

The helium abundance, $Y_{\rm He}$, can be measured by the cosmic microwave background (CMB) since the damped tail of the CMB anisotropies is affected by the free electron density between the helium and hydrogen recombination, which is modified by variations in \mbox{$Y_{\rm He}$ \citep{2016A&A...594A..13P}}. Using this method and allowing $Y_{\rm He}$ to vary as a derived parameter in the framework of $\Lambda$CDM model, \citet{2020A&A...641A...6P} gave the following constraints from the Planck TT, TE, EE, and lowE datasets at a 95\% confidence level:
\begin{equation}\label{planckYhe}
Y_{\rm He}=0.241 \pm 0.025~.
\end{equation}%MDPI: Please check if indenting is required here. Similar highlights are same meaning.
In addition to the CMB anisotropies, the helium abundance can also affect the stellar evolution and galactic chemical evolution. 
Traditional measurements of the helium abundance are mainly based on the big bang nucleosynthesis (BBN) theory \citep{2015JCAP...07..011A,2016RMxAA..52..419P,2014MNRAS.445..778I}. In this paper, we take the results of \citet{2015JCAP...07..011A} as a reference, giving a slightly tighter constraint
\begin{equation}
Y_{\mathrm{P}}^{\mathrm{BBN}}=0.2449 \pm 0.0040(68 \% \text { CL}).
\end{equation}

Several approaches also have been employed to constrain the helium abundance, by using the integrated spectra to infer the helium abundance of extragalactic globular \mbox{clusters \citep{2022MNRAS.tmp..584L}}, via observations of metal poor HII regions to determine the primordial Helium \mbox{abundance \citep{2013JCAP...11..017A},} by observations of the Extremely Metal-Poor Galaxies to determine the primordial helium abundance \citep{2022arXiv220309617M}, and so on. However, these methods suffer from several uncontrolled systematic errors.

%A reasonable way to address them is to give model-independent constraints on helium abundance to test the results.

%As we know, the electron fraction $\chi_e$ is directly related to the primordial hydrogen and helium abundance $Y_{\rm H}$ and $Y_{\rm He}$. Through the electron fraction, we can see that the DM of FRBs and $Y_{\rm He}$ is strongly connected. Therefore, if we could measure the DM of FRBs and build the relationship between DM and redshift, the Helium abundance could also be measured by the FRBs with less systematic effect.
%Furthermore, the connection between DM and $\chi_e$ is very clear. So we can establish a connection between DM and $Y_{He}$. With ever-increasing samples, FRBs will be powerful probes for limiting helium abundance.
%In this paper, we study the measurement of the helium abundance via DM - z relation using localized FRBs samples. We introduce the methods for probing cosmological parameters through DM - z relation in Section \ref{sec:DM}. In Section \ref{Results}, we present the limited results for 17 FRB samples that have been localized so far. In Section \ref{Future Prospects}, we explore the possibilities of helium abundance measurements by simulating FRBs samples which can be available in the future. Finally, we present conclusions in Section \ref{Conclusion}.
It is well known that the electron fraction, $\chi_e$, is directly related to the original hydrogen and helium abundances $Y_{\rm H}$ and $Y_{\rm He}$. Using the electron fraction, we can see that there is a strong degeneracy between the DM of the FRB and $Y_{\rm He}$. Therefore, if we can measure the DM of FRBs and establish a relationship between the DM and redshift and, in the meantime, use observations to precisely constrain other cosmological parameters, the helium abundance can also be measured by FRBs, avoiding large systematic errors. In this paper, we investigate the measurement of helium abundance using current and future FRB samples. We present methods for probing cosmological parameters via the DM-$z$ relationship in Section \ref{sec:DM}. In Section \ref{Results}, we present the constraint results on $Y_{\rm He}$ from 17 current \mbox{FRB} samples. In Section \ref{Future Prospects}, by introducing the shift parameters of the Planck measurements, we explore the possibilities of the helium abundance measurements by simulating future FRB samples. Finally, we present our conclusions in Section \ref{conclusion}.

%%%%%%%%%%%%%%%%%%%%%%%%%%%%%%%%%%%%%%%%%%
\section{Properties of FRBs} \label{sec:DM}

%The total observed DM$_{obs}$ consists of three components~\citep{2013Sci...341...53T,2014ApJ...783L..35D}:
%\begin{equation}
%\mathrm{DM}_{\mathrm{obs}}=\mathrm{DM}_{\mathrm{host}}+\mathrm{DM}_{\mathrm{IGM}}+\mathrm{DM}_{\mathrm{MW}}
%\end{equation}
%where DM with different subscripts denote contribution from the host galaxy, the intergalactic medium(IGM) and the Milky Way, respectively.

The observed dispersion measure, ${\rm DM_{obs}}$, is defined as the integral of the free electrons number density along the line of sight, which consists of the contributions from the IGM, ${\rm DM_{IGM}}$; the FRB host galaxy, ${\rm DM_{host}}$; and the Milky Way, ${\rm DM_{MW}}$ \citep{2013Sci...341...53T,2014ApJ...783L..35D}. ${\rm DM_{IGM}}$ from a fixed source redshift $z$ is given by:
\begin{equation}
    {\rm DM_{IGM}}\left(z\right)=\int^{z}_{0}\frac{{\rm d}z}{H(z)}\frac{n_e(z)f_{\rm IGM}(z)}{(1+z)^2}~,
\end{equation}
where $\Omega_{b}$ represents the baryon density parameter, $E(z)$ is the dimensionless expansion function $E(z)=H(z)/H_0$, and $f_{IGM}(z)$ represents the fraction of electrons in the IGM.
Since we only consider the FRB sample with $z<3$ and assume that both hydrogen and helium are fully ionized, the cosmic electron density can be expressed as a function of the baryon abundance, $n_e(z)=\rho_b(z)\chi_e(z)/m_{\rm p}$. Here, $\rho_b(z)$ is the baryon mass density, $m_{\rm p}$ is the proton mass, and the electron fraction is :
\begin{equation}
    \chi_e(z)=Y_{\rm H}+\frac{1}{2}Y_{\rm He}\approx\left(1-Y_{\rm He}\right)+\frac{1}{2}Y_{\rm He}=1-\frac{1}{2}Y_{\rm He}~,
\end{equation}
which is related to the primordial hydrogen and helium abundances. At present the CMB measurement provides the constraint on the helium abundance, $Y_{\rm He}=0.241\pm0.025$ \citep{2020A&A...641A...6P}. Since there is little star formation at low redshifts, the overall fraction of electrons in the IGM does not evolve significantly over the redshift range covered by the FRB \mbox{sample \citep{2019ApJ...876..146L,Dai:2021czy},} and we simply keep $f_{\rm IGM}=0.84$ constant.

Finally, we can write the relation between the DM and redshifts as:
\begin{equation}\label{IGMequation}
{\rm DM_{IGM}}(z)=\frac{3 c H_{0} \Omega_{b} f_{\rm IGM}}{8 \pi G m_{\rm p}} \left(1-\frac{1}{2}Y_{\rm He}\right) \int_{0}^{z} \frac{\left(1+z^{\prime}\right) {\rm d} z^{\prime}}{E\left(z^{\prime}\right)}~,
\end{equation}
where $c$ represents the speed of light, and $G$ is the gravitational constant.

The distribution of electrons in the IGM is inhomogeneous, and there is a stochastic contribution to the dispersion measure of a large scale structure, both of which lead to the complicated uncertainty of ${\rm DM_{IGM}}$. In this paper, for simplicity, we consider a Gaussian distribution around the mean value of ${\rm DM_{IGM}}$, and we interpolate the standard deviation linearly from the values found in simulations, using $\sigma_{\rm IGM}(z=0)\approx40\,{\rm pc\,cm^{-3}}$ and $\sigma_{\rm IGM}(z=1)\approx180\,{\rm pc\,cm^{-3}}$. Due to the lack of understanding of the high redshift universe, we 
naively extend this relation, $\sigma_{\rm IGM}(z)\approx40+140z\,{\rm pc\,cm^{-3}}$, to the high redshift, which is roughly similar to the numerical simulation results \citep{2014ApJ...780L..33M} at $z\sim1.5$. Since most of the FRB samples we have detected and the simulated samples that appear below lay at redshifts $z<2$, the linear relationship we assume for $\sigma_{IGM}$ affects the results very little.

For typical FRBs, there are two objects along the line of sight: the host halo and the Milky Way. The Milky Way DM can be predicted and removed with the help of models of the galactic electron distribution. We use the NE2001 model \citep{2002astro.ph..7156C} to subtract the Milky Way contribution for each FRB position in the sky. For sources at high galactic latitude $\left(|b|>10^{\circ}\right)$ where most FRBs are detected, the average uncertainty of the DM contribution from the Milky Way, $\sigma_{\mathrm{MW}}$, is about 30 ${\rm pc\,cm^{-3}}$ \citep{2005AJ....129.1993M}; therefore, we take $\sigma_{\rm MW}\approx30\,{\rm pc\,cm^{-3}}$ as a measure for the uncertainty of the model.

The host galaxy properties are more uncertain, due to the dependence on the type of the host galaxy, the relative orientations, and the near-source plasma, which are poorly known. \citet{2020Natur.581..391M} estimated ${\rm DM_{host}}\approx\,50/(1+z_{\rm host})\,{\rm pc\,cm^{-3}}$ theoretically from the localized FRBs. {However, in our analysis, we further assume that the host halos are more or less similar to the Milky Way, ${\rm DM_{host}}\approx\,100/(1+z_{\rm host})\,{\rm pc\,cm^{-3}}$, conservatively, due to the large value of DM$_{host}\in[55,\,225]$ $\rm pc\,cm^{-3}$ of FRB 121102}, and allow for a large scatter $\sigma_{\rm{host}}\approx\,50/(1+z_{\rm host})\,{\rm pc\,cm^{-3}}$ \citep{2003ApJ...598L..79I,2014ApJ...783L..35D}.

Of the hundreds of verified FRBs publicly available, only 19 FRBs have been localized at present, including the nearest repeating FRB 200110E \citep{2022Natur.602..585K} and FRB 181030A \citep{2021ApJ...919L..24B}. In our analysis, we mainly used the 17 localized FRBs\endnote{\url{http://frbhosts.org} (4th June,2022).}, which are listed in Table \ref{tab:17FRBs}, to perform the numerical constraints and neglected the nearest FRB 200110E and FRB 181030A.%MDPI: Please add accessed date (Date Month Year, the exact date when you last accessed the link). 

\vspace{-6pt}%{}
\begin{table}[H]
\centering
\caption{Overview of all 17 FRBs with {the measured redshift}. {The Milky Way DM is predicted by using the NE2001 model}.}
\label{tab:17FRBs}

\begin{adjustwidth}{-\extralength}{0cm}
\newcolumntype{C}{>{\centering\arraybackslash}X}
\newcolumntype{B}{>{\raggedright\arraybackslash}X}
\newcolumntype{A}{>{\raggedleft\arraybackslash}X}
%\begin{tabularx}{\textwidth}{CCC}
\begin{tabularx}{\fulllength}{CCCCCm{5cm}<{\centering}
}

\toprule
 \textbf{Name} & \textbf{Redshift} & \boldmath{$\rm DM_{obs}$} \boldmath{$({\rm pc\,cm^{-3}})$} & \boldmath{$\rm DM_{MW}$} \boldmath{$({\rm pc\,cm^{-3}})$} & \textbf{Telescope} & \textbf{Reference} \\
\midrule
 FRB 121102 & $0.19273$ & 557 & $188.0$ & Arecibo & \citet{2017Natur.541...58C} \\
FRB 180916 & $0.0337$ & $348.8$ & $200.0$ & CHIME & \citet{2020Natur.577..190M} \\
FRB 180924 & $0.3214$ & $361.42$ & $40.5$ & ASKAP & \citet{2019Sci...365..565B} \\
FRB 181112 & $0.4755$ & $589.27$ & $102.0$ & ASKAP & \citet{2019Sci...366..231P} \\
FRB 190102 & $0.291$ & $363.6$ & $57.3$ & ASKAP & \citet{2020ApJ...895L..37B} \\
FRB 190523 & $0.66$ & $760.8$ & $37.0$ & DSA-10 & \citet{2019Natur.572..352R,2020ApJ...903..152H} \\
FRB 190608 & $0.1178$ & $338.7$ & $37.2$ & ASKAP & \citet{2021ApJ...922..173C} \\
FRB 190611 & $0.378$ & $321.4$ & $57.8$ & ASKAP & \citet{2020ApJ...903..152H}\\
FRB 190614 & $0.6$ & $959.2$ & $83.5$ & VLA & \citet{2020ApJ...899..161L} \\
FRB 190711 & $0.522$ & $593.1$ & $56.4$ & ASKAP & \citet{2020ApJ...903..152H} \\
FRB 190714 & $0.2365$ & $504.13$ & $38.0$ & ASKAP & \citet{2020ApJ...903..152H}) \\
FRB 191001 & $0.234$ & $507.9$ & $44.7$ & ASKAP & \citet{2020ApJ...903..152H} \\
FRB 200430 & $0.16$ & $380.25$ & $27.0$ & ASKAP & \citet{2020ApJ...903..152H} \\
FRB 201124 & $0.098$ & $413.52$ & $123.2$ & ASKAP & \citet{2021ATel14515....1D,2021arXiv210609710R} \\
FRB 180301 & $0.3304$ & $536$ & $ 152$ & Parkes & \citet{2022AJ....163...69B} \\
FRB 191228 & $ 0.2432$ & $297.5$ & $ 33$ & ASKAP & \citet{2022AJ....163...69B} \\
FRB 200906 & $ 0.3688$ & $ 577.8$ & $ 36$ & ASKAP & \citet{2022AJ....163...69B} \\
\bottomrule

\end{tabularx}
\end{adjustwidth}
\end{table}

\section{Constraints from Current Data} \label{Results}

Assuming the flat $\Lambda$CDM model, in our calculations, we performed a global fitting analysis using the public {\tt CosmoMC} software package \citep{PhysRevD.66.103511}, to constrain three  parameters: the helium abundance $Y_{\rm He}$, the Hubble constant $H_0$, and the dimensionless baryon density $\Omega_b$. 
Since we are studying the constraints on the helium abundance from FRBs, we set the choice of not using BBN consistency in our analysis, and our most general parameter space was:
\begin{equation}\label{parameter space}
\mathbf{P} \equiv\left(\Omega_{b} h^{2}, \Omega_{c} h^{2}, \Theta_{s}, \tau, Y_{He}, n_{s}, \ln \left( A_{s}\right)\right)
\end{equation}
where $\Omega_{b}$ and $\Omega_{c}$ are the baryon and cold dark matter densities relative to the critical density, $\Theta_{s}$ is the ratio (multiplied by 100) of the sound horizon at decoupling to the angular diameter distance to the last scattering surface, $\tau$ is the optical depth to reionization, $Y_{He}$ is the helium abundance, and $A_{s}$ and $n_{s}$ are the amplitude and the tilt of the power spectrum of the primordial scalar perturbations, respectively.
Here, we set $\Omega_{c} h^{2} = 0.1202$, $\tau = 0.0544$, $n_{s} = 0.96$, and $ln\left( A_{s}\right) = 3.1$ as fixed values and varied $\Omega_{b} h^{2}$, $\Theta_{s}$, and $Y_{He}$ in our analysis.
%Here we fix $\Omega_{c} h^{2} = 0.1202$, $\tau = 0.0544$, $n_{s} = 0.96$, $ln\left( A_{s}\right) = 3.1$ for fixed value, and vary $\Omega_{b} h^{2}\in[0.005,\,0.1]$, $\Theta_{s}\in[0.5,\,2]$, $Y_{He}\in[0.1,\,0.5]$ in our analysis. }

For the data analysis, we assumed Gaussian individual likelihoods to observe a dispersion measure ${\rm DM}_i$ at a given redshift $z_i$:
\begin{equation}\label{likelihood}
\mathcal{L}\left(\mathrm{DM}_{i}, z_{i}\right)=\frac{1}{\sqrt{2 \pi \sigma_{i}^{2}}} \exp \left(-\frac{\left(\mathrm{DM}_{i}-\mathrm{DM}_{\rm{IGM}}\left(z_{i}\right)\right)^{2}}{2 \sigma_{i}^{2}}\right)~.
\end{equation}
The total variance for the DM measurement of each FRB follows from the individual uncertainties accounting for the scatter of the IGM contribution, the MW electron distribution model, and the host galaxy:
\begin{equation}
\sigma_{i}^{2}(z_{i})=\sigma_{\mathrm{MW}}^{2}+\sigma_{\rm{host }}^{2}(z_{i})+\sigma_{\mathrm{IGM}}^{2}(z_{i}),
\end{equation}
Apparently, at high redshifts, the uncertainty of ${\rm DM_{IGM}}$ will dominate the whole variance of the DM.

\begin{figure}[H]
	%\centering
    \includegraphics[width=0.8\linewidth]{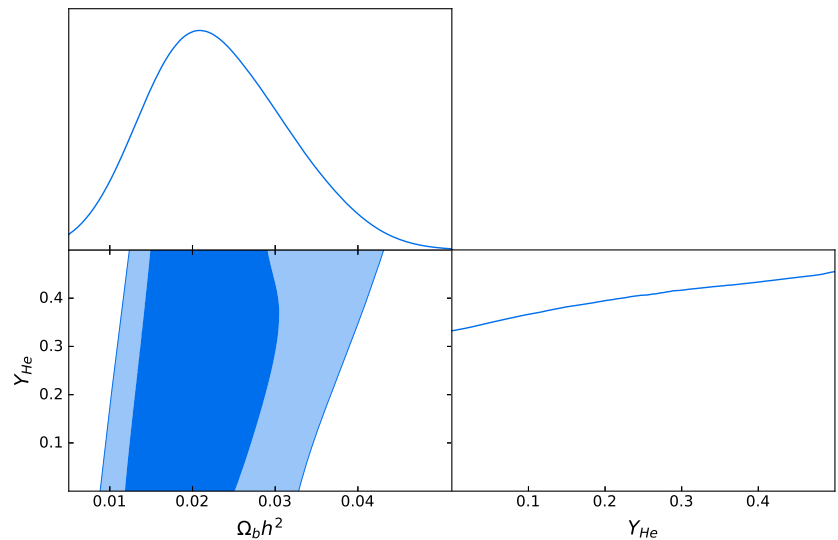}
    \vspace{-6pt}%{}
	\caption{One-dimensional and two-dimensional constraints on the $Y_{\rm He}$ and $\Omega_bh^2$ from the 17 current FRBs.}
	\label{fig:17realdata}
\end{figure}

In Figure~\ref{fig:17realdata}, we see that the current 17 FRB data had very limited constraining ability for the helium abundance, whose posterior distribution was almost flat. The main reason is that the redshifts of these 17 FRBs are relatively low, while the helium abundance is associated with the universe at higher redshifts. Therefore, it is impossible to understand cosmological information at high redshifts through these FRBs with low redshifts. Second, in Figure~\ref{fig:17realdata}, we also show the constraint on the baryon energy density $\Omega_bh^2$, which was also very weak compared to other observations. According to Equation~(\ref{IGMequation}), there is a strong degeneracy between $Y_{\rm He}$ and $\Omega_bh^2$, which was also confirmed in the two-dimensional constraint of Figure~\ref{fig:17realdata}. Therefore, it is very difficult to obtain useful results only from the  \mbox{17 current FRBs' data.}

%For Gaussian distributed measurements, the likelihood function $L \propto e^{-\chi^{2} / 2}$, with
%\begin{equation}
%\chi_{DM}^{2}=\left[\frac{\left(DM_i-DM_{IGM}(z_i)\right)}{\sigma_{all}}\right]^{2}
%\end{equation}
%and the complete expression is
%\begin{equation}
%\mathcal{L}\left(\mathrm{DM}_{i}, z_{i}\right)=\frac{1}{\sqrt{2 \pi \sigma_{all}^{2}}} \exp \left(-\frac{\left(\mathrm{DM}_{i}-\mathrm{DM}_{\text {IGM }}\left(z_{i}\right)\right)^{2}}{2 \sigma_{all}^{2}}\right).
%\end{equation}

%We use the public COSMOMC package to perform a global fitting analysis~\citep{PhysRevD.66.103511}.

%From the figure we can see that the constraints are not good, mainly due to the lack of samples and the degeneracy between helium abundance and $\Omega_bh$.

\section{Future Prospects} \label{Future Prospects}

The current limited FRB data can not provide precise constraints on the helium abundance. Fortunately, the amount of available FRBs is expected to grow quickly over the next few years. In this section, we investigate the constraining ability of FRBs on the helium abundance from a future mock sample.

\subsection{Mock Data} \label{Mock data}

\begin{figure}[H]
\hspace{-15pt}%{}
\begin{tabular}{cc}
\includegraphics[width=7cm]{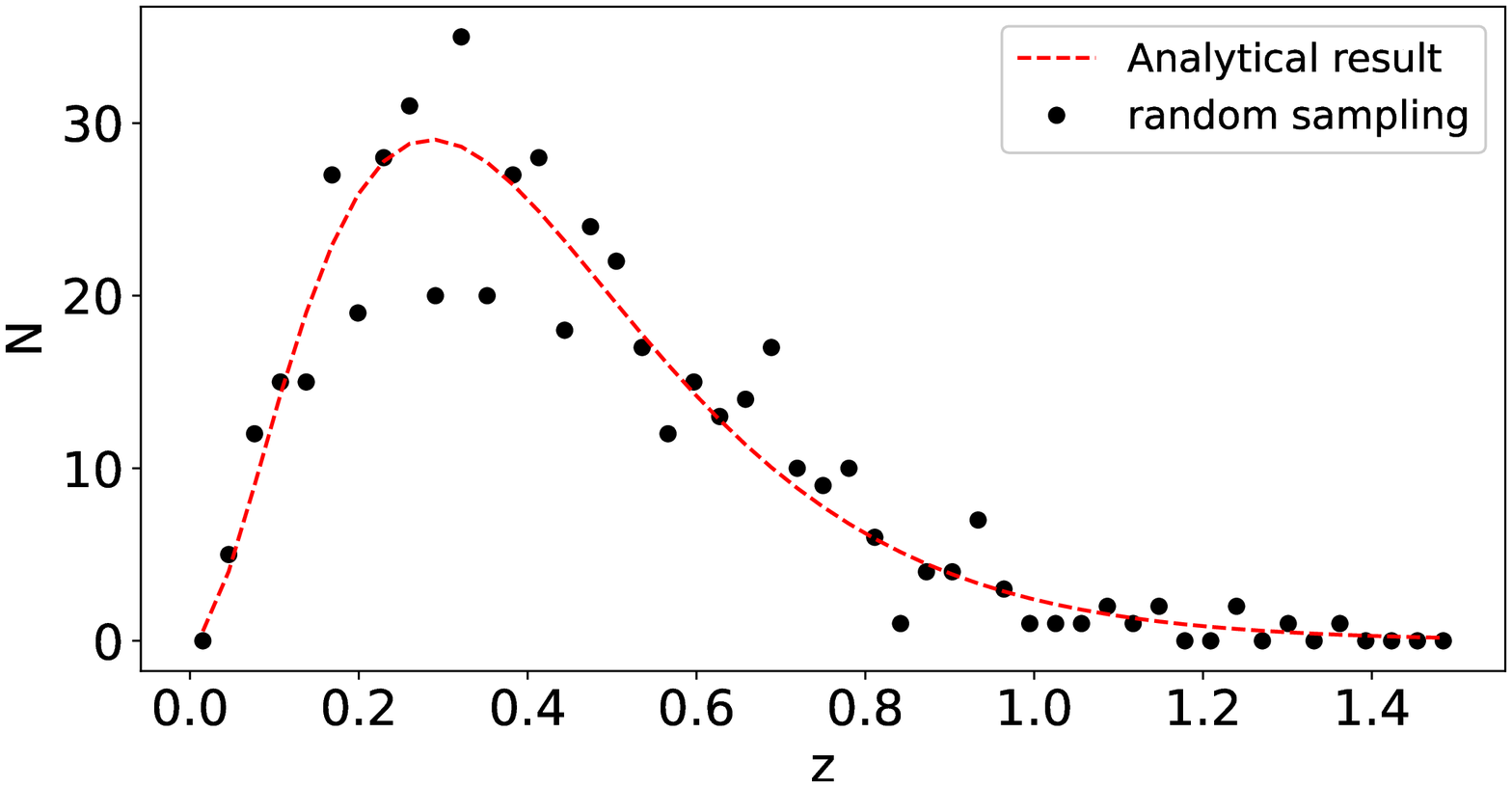} &
 \includegraphics[width=7cm]{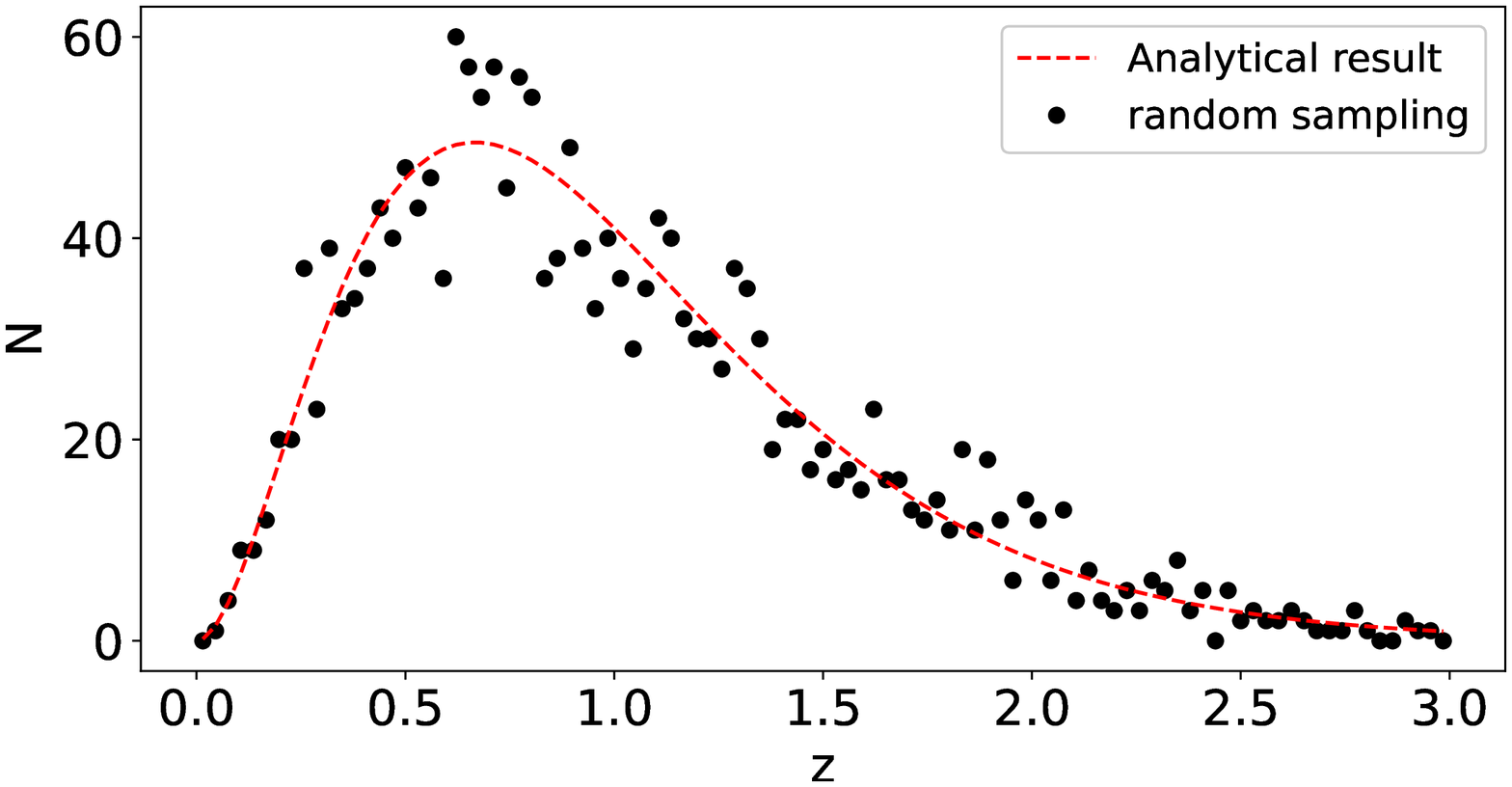} \\
(\textbf{a}) & (\textbf{b}) \\
\end{tabular}
\caption{Redshift distributions of the simulated FRBs samples. (\textbf{a}) Conservative case: 500 FRBs with $\alpha = 7$. (\textbf{b}) High-redshift case: 2000 FRBs with $\alpha = 3$.}
\label{fig:redshiftdistribution}
\end{figure}

\citet{2021PhRvD.103h3536Q} studied the effect of the FRB redshift distribution on cosmological constraints in detail. However, the current limited samples can not provide us with the accurate information of FRBs' redshift distribution. For simplicity, we generated the mock data from the FRB redshift distribution following the galaxy distribution, which can be written as:
\begin{equation}
n(z)=z^{2} \exp (-\alpha z)~,
\end{equation}
where $\alpha$ denotes the effective depth of the sample. Considering that the majority of FRB detection lies most likely at lower redshifts $z<1$, we firstly created a sharp cutoff with $\alpha = 7$ and, conservativelym generated 500 samples up to $z\sim 1.5$ (the conservative case). On the other hand, since the future SKA measurement has sufficient sensitivity to detect high-redshift FRBs, we also set $\alpha = 3$ to generate 2000 samples up to $z\sim 3$ for comparison (the high-redshift case). The fiducial values of the related parameters were: $\Omega_bh^2 = 0.02230$, $h = 0.671$, $\Omega_m = 0.318$, $Y_{\rm He} = 0.24$, and $f_{\rm IGM}=0.84$ to generate ${\rm DM_{IGM}}$, and the obtained sampling results are shown in Figure~\ref{fig:redshiftdistribution} for these two cases.

\vspace{-6pt}%{}

%Considering that there is no evidence for redshift evolution of $f_{IGM}$, we fix the value of $f_{IGM}$ to 0.84 for simplicity. The total variance still uses the variance in \ref{sec:DM}. We set $\Omega_bh^2 = 0.02235$, $h = 0.7$, $\Omega_m = 0.3$, $Y_{He} = 0.24$ to generate DM$_{IGM}$ from redshift samples.
%The corresponding results are shown in Figure \ref{fig:results_withoutPrior}

\begin{figure}[H]
	%\centering
    \includegraphics[width=0.8\linewidth]{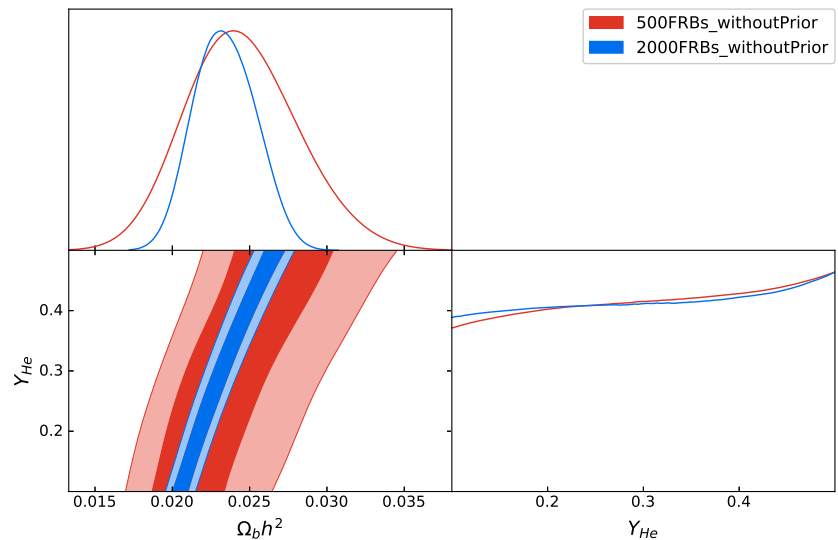}\vspace{-6pt}%{}
	\caption{One-dimensional and two-dimensional constraints on the $Y_{\rm He}$ and $\Omega_bh^2$ from the conservative case with $N=500$ (red) and the high-redshift case with $N=2000$ (blue).}
	\label{fig:results_withoutPrior}
\end{figure}

Extrapolating from the current uncertainty of the measurement, the constraint on the baryon energy density $\Omega_bh^2$ was improved by a factor of 2, when considering the conservative case with $N=500$. If we further consider the higher redshift case with $N=2000$ FRB data, the limit on $\Omega_bh^2$ would further improved, and the standard deviation is about 
$0.002$. However, as can be seen in Figure~\ref{fig:results_withoutPrior}, unfortunately, no matter what kind of mock data we consider, it is still impossible to give a reasonably restricted result for the helium abundance. This is because although the number of mock data has enlarged, if we cannot provide better constraints on $\Omega_bh^2$ by using other independent observations, the strong degeneracy between the helium abundance and the baryon energy density still cannot be broken, and the constraints on the helium abundance also will not be improved.

\vspace{-6pt}%{}
%We can see from the figure that the result still does not converge, since there is a strong degeneracy between $\Omega_bh$ and $Y_{He}$, we need to introduce an external method to break the degeneracy.

\subsection{Shift Parameters} \label{Shift Parameters}

There are several studies on the degeneracies between cosmological parameters~\mbox{\citep{2013MNRAS.434.1792S,2011ApJ...743...28K,2014JCAP...02..012C},} and one way to break the degeneracy between $Y_{\rm He}$ and $\Omega_bh^2$ is usually to use a Gaussian prior on $\Omega_bh^2$, as given by \citet{2018ApJ...855..102C}. 
In this paper, we introduce the shift parameters from the CMB measurements, which provides partial information of the CMB anisotropies, especially the distance information, which can provide constraints of cosmological parameters to some extent without using the CMB full power spectra \citep{2007PhRvD..76j3533W,2008ApJ...683L...1L}.  {When compared with the CMB full data, we can obtain similar results on parameters by using the shift parameter method, without consuming too much time.} 
In practice, we firstly used the full CMB power spectrum from Planck to obtain the best-fit values and the inverse covariance matrix of these shift parameters, and then input them into our calculations as a prior to break the degeneracy. 
It is worth noting that breaking the degeneracy with a Gaussian prior gives about the same effect; we introduced the method of shift parameters, since our calculations of the inverse covariance matrix among these shift parameters can also be applied in constraining other models.

The shift parameters, $R$ and $l_A$, correspond to the scaled distance to recombination and the angular scale of the sound horizon at recombination, respectively, given by
\begin{equation}
R\left(z_{*}\right) =\sqrt{\Omega_{m} H_{0}^{2}} \chi\left(z_{*}\right) 
\end{equation}
\begin{equation}
l_{A}\left(z_{*}\right) =\pi \chi\left(z_{*}\right) / \chi_{s}\left(z_{*}\right)
\end{equation}
where $\chi\left(z_{*}\right)$ denotes the comoving distance to $z_{*}$, and $\chi_{s}\left(z_{*}\right)$ denotes the comoving sound horizon at $z_{*}$. Furthermore, the decoupling epoch, $z_{*}$, is given by~\citep{1998ApJ...496..605E}
\begin{equation}
z_{*}=1048\left[1+0.00124\left(\Omega_{b} h^{2}\right)^{-0.738}\right]\left[1+g_{1}\left(\Omega_{m} h^{2}\right)^{g_{2}}\right]
\end{equation}
where
\begin{equation}
g_{1}=\frac{0.0783\left(\Omega_{b} h^{2}\right)^{-0.238}}{1+39.5\left(\Omega_{b} h^{2}\right)^{0.763}}, \quad g_{2}=\frac{0.560}{1+21.1\left(\Omega_{b} h^{2}\right)^{1.81}}.
\end{equation}
The comoving sound horizon $\chi_{s}\left(z_{*}\right)$ is given by
$$
\chi_{s}(z)=\frac{c}{\sqrt{3}} \int_{0}^{1 /(1+z)} \frac{d a}{a^{2} H(a) \sqrt{1+\left(3 \Omega_{b} / 4 \Omega_{\gamma}\right) a}},
$$
where $\Omega_{\gamma}=2.469 \times 10^{-5} h^{-2}$ for $T_{\mathrm{cmb}}=2.725 \mathrm{~K}$, and
\begin{equation}
H(a)=H_{0}\left[\frac{\Omega_{m}}{a^{3}}+\frac{\Omega_{r}}{a^{4}}+\Omega_{\Lambda}\right]^{1 / 2} ,
\end{equation}
where $\Omega_{\Lambda}$ is the dark energy density parameter, and the radiation density parameter, $\Omega_{r}$, is the sum of photons and relativistic neutrinos,
\begin{equation}
\Omega_{r}=\Omega_{\gamma}\left(1+0.2271 N_{\text {eff }}\right),
\end{equation}
where $N_{\rm eff} = 3.04$ is the effective number of neutrino species.

We performed a global fitting analysis using the {\tt CosmoMC} package to constrain $R$, $l_A$, and $\Omega_bh^2$ from the CMB power spectrum of the Planck measurement in the framework of the standard $\Lambda$CDM model.
Our most general parameter space follows Equation~\eqref{parameter space}, fixes $Y_{He} = 0.24$, and varies $\Omega_{b} h^{2}$, $\Omega_{c} h^{2} $, $\Theta_{s}$, $\tau$, $n_{s}$, and $ln\left( A_{s}\right)$ in our analysis.  
%\added{Our most general parameter space still follows Equation~\ref{parameter space}, and vary $\Omega_{b} h^{2}\in[0.005,\,0.1]$, $\Omega_{c} h^{2} \in[0.001,\,0.99]$, $\Theta_{s}\in[0.5,\,10]$, $\tau\in[0.01,\,0.8]$, $n_{s}\in[0.8,\,1.2]$, $ln\left( A_{s}\right)\in[2,\,4]$, $Y_{He}\in[0.1,\,0.5]$ in our analysis.} 
In Table~\ref{tab:matrix}, we list the mean values of $R$, $l_A$, and $\Omega_bh^2$ and the inverse covariance matrix among them. It is worth noting that the matrix we used was based on a fixed $Y_{He} = 0.24$, and the result did not change significantly if we vary $Y_{He}$.

Since we put the information of the shift parameters into the calculations, which is independent on the DM of the FRBs, the joint likelihood of the sample is then the product of the individual likelihoods, and the corresponding $\chi^2$ function becomes:
\begin{equation}
\chi^{2}=\chi_{\rm DM}^{2}+\chi_{\rm shift}^{2},    
\end{equation}
where the $\chi^2$ function of shift parameters can be written as:
\begin{equation}
\chi_{\rm shift}^{2} =\left(x_{i}^{\rm th}-x_{i}^{\rm data}\right)\left(C_{ij}^{-1}\right)\left(x_{j}^{\rm th}-x_{j}^{\rm data}\right)
\end{equation}
where $x=\left(\Omega_bh^2, R, l_{A}\right)$ is the parameter vector, and $\left(C_{ij}^{-1}\right)$ is the inverse covariance matrix among these shift parameters, which is shown in Table~\ref{tab:matrix}.
{In our analysis, we adopted this updated likelihood function, which included the information of $\Omega_bh^2$, to break the degeneracy between $Y_{He}$ and $\Omega_bh^2$.}

\begin{table}[H]
	%\centering
	\caption{The mean values of shift parameters: $l_{A}, R,$ and $\Omega_bh^2$ from the Planck measurement and the corresponding inverse covariance matrix among them.}
	\label{tab:matrix}
\newcolumntype{C}{>{\centering\arraybackslash}X}
\newcolumntype{B}{>{\raggedright\arraybackslash}X}
\newcolumntype{A}{>{\raggedleft\arraybackslash}X}
\begin{tabularx}{\textwidth}{CCCCC}
%\begin{tabularx}{\fulllength}{CCCCCC}

\toprule
 & \boldmath{$l_{A}\left(z_{*}\right)$} & \boldmath{$R\left(z_{*}\right)$} & \boldmath{$\Omega_bh^2$} \\
\midrule%MDPI: Please use Scientific notation. please revise all text.
$l_{A}\left(z_{*}\right)$ & $1.169\times10^2$ & $-1.164\times10^3$ & $5.325\times10^2$ \\
$R\left(z_{*}\right)$ &$-1.164\times10^3$ & $9.236\times10^4$ & $1.631\times10^6$ \\
$\Omega_bh^2$ &$5.325\times10^2$ & $1.631\times10^6$ & $7.844\times10^7$ \\
\midrule

Mean Value & $3.018\times10^2$ & $1.750\times10^0$ & $2.230\times10^{-2}$\\
\bottomrule

\end{tabularx}
\end{table}

\subsection{Results from DM and Shift Parameters} \label{Results of mock data with prior}

%Since we get the shift parameters from the Planck data, we need to simulate the FRB data using the optimal values of the parameters given by the Planck data, where set $\Omega_bh^2 = 0.02230$, $h = 0.671$, $\Omega_m = 0.318$, $Y_{He} = 0.24$ to generate DM$_{IGM}$ from redshift samples. After breaking the degeneracy using the shift parameter as a prior, we obtain the final helium abundance constraints from the 500 and 2000 simulated data, as shown in Figure \ref{fig:final results}.

\begin{figure}[H]
%	\centering
    \includegraphics[width=0.6\linewidth]{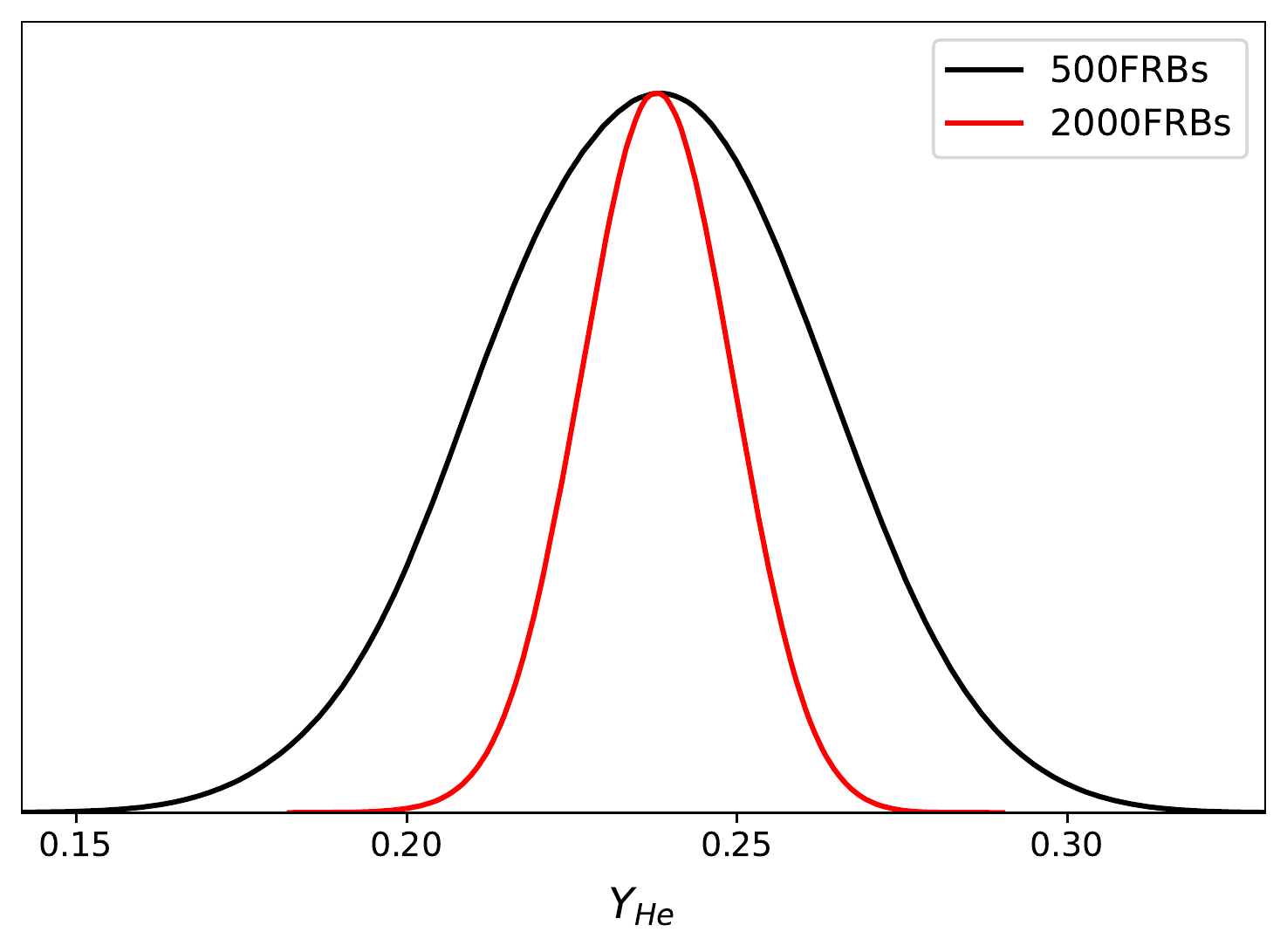}
	\caption{One-dimensional constraints on the helium abundance from the conservative case (black line) and the high-redshift case (red line), respectively.}
	\label{fig:final results}
\end{figure}

Firstly, we considered the conservative case with $N=500$ mock data. In Figure~\ref{fig:final results}, we present the one-dimensional constraint on the helium abundance from the FRB DM data, together with the prior of shift parameters. We can clearly see that, due to the tight constraint on the baryon energy density, $\sigma(\Omega_bh^2)=0.0001$, the degeneracy between $Y_{\rm He}$ and $\Omega_bh^2$ has been broken entirely. The consequent constraint on the helium abundance becomes reasonable with the standard deviation $\sigma(Y_{\rm He})=0.025$, instead of an almost flat distribution, as shown in Figure~\ref{fig:results_withoutPrior}. Furthermore, this constraint is not much different from the current CMB constraint $Y_{\rm He}=0.241 \pm 0.025$ at a 95\% confidence level and BBN constraint $\sigma(Y_{\mathrm{P}}^{\mathrm{BBN}})=0.004$. This result indicates that with the help of external shift parameters information, we can use the DM of FRBs mock data to provide a good constraint on the \mbox{helium abundance.}

Then, we moved to the high-redshift case with more FRB data $N=2000$ and a higher redshift distribution up to $z\sim3$. Since the obtained error bar on the parameters should be smaller with $N^{-1/2}$, we expect that the constraint on the helium abundance would be improved by a factor of 2. In practice, we used the 2000 mock FRBs and the shift parameters to perform the analysis and obtain the standard deviation of the helium abundance $\sigma(Y_{\rm He})=0.011$, which was slightly better than we expected, as shown in the red line of Figure~\ref{fig:final results}. We think that this is due to the redshift distribution of these mock data, which was higher than the conservative case. We could have more information from the high redshift universe to study the helium abundance, which would be helpful to improve the constraint on $Y_{\rm He}$. In the future, the SKA project expects to detect FRB samples with redshifts up to 14 \citep{2016JCAP...05..004F}, which implies FRB could be a very promising probe to study the high-redshift universe.

\section{Conclusions} \label{conclusion}

In this paper, we performed an analysis to study the helium abundance by using current and future FRB data, based on the DM-$z$ relation. Here, we summarize our \mbox{main conclusions:}
\begin{itemize}
    \item Since the 17 current FRB samples have low redshift, resulting in poor quality of the samples, we could not obtain a useful constraint on the helium abundance, which is associated with the universe at higher redshifts.
    \item Then we simulated two mock data: the conservative case at low redshift and the high-redshift case. However, due to the strong degeneracy between the helium abundance and the baryon energy density, the constraints on $Y_{\rm He}$ were still very weak from the mock FRB data.
    \item Therefore, we introduced the distance information of shift parameters, derived from the CMB full power spectra of the Planck measurement. With this help, the constraint on the baryon energy density was significantly improved, and the degeneracy with $Y_{\rm He}$ was broken. 
    \item Consequently, the constraints on the helium abundance were also improved with the standard deviation $\sigma(Y_{\rm He})=0.025$ and $0.011$ for two FRBs' mock data, respectively. As can be seen from the current CMB constraint $Y_{\rm He}=0.241 \pm 0.025$ at a 95\% confidence level and BBN constraint $\sigma(Y_{\mathrm{P}}^{\mathrm{BBN}})=0.004$, the constraints from the FRBs are comparable.  Hopefully,  large FRB samples with high redshift from the Square Kilometre Array will provide high-precision measurements of the helium abundance in the near future.
\end{itemize}

\vspace{6pt} 

%%%%%%%%%%%%%%%%%%%%%%%%%%%%%%%%%%%%%%%%%%
%% optional
%\supplementary{The following are available online at \linksupplementary{s1}, Figure S1: title, Table S1: title, Video S1: title.}


%%%%%%%%%%%%%%%%%%%%%%%%%%%%%%%%%%%%%%%%%%
\authorcontributions{Formal analysis, J.-Q. X.; investigation, L.J.; Writing---original draft preparation, L.J.; Writing---review and editing, J.-Q.X. All authors have read and agreed to the published version of the manuscript.}%For research articles with several authors, a short paragraph specifying their individual contributions must be provided. The following statements should be used ``Conceptualization, X.X. and Y.Y.; methodology, X.X.; software, X.X.; validation, X.X., Y.Y. and Z.Z.; formal analysis, X.X.; investigation, X.X.; resources, X.X.; data curation, X.X.; writing---original draft preparation, X.X.; writing---review and editing, X.X.; visualization, X.X.; supervision, X.X.; project administration, X.X.; funding acquisition, Y.Y. All authors have read and agreed to the published version of the manuscript.'', please turn to the \href{http://img.mdpi.org/data/contributor-role-instruction.pdf}{CRediT taxonomy} for the term explanation. Authorship must be limited to those who have contributed substantially to the work~reported.

\funding{J.-Q. Xia is supported by the National Science Foundation of China under grants No. U1931202 and 12021003; the National Key R\&D Program of China No. 2020YFC2201603.}

\dataavailability{Data relevant to this study are available on request to the corresponding~\mbox{author.}}

\institutionalreview{Not applicable}%In this section, please add the Institutional Review Board Statement and approval number for studies involving humans or animals. Please note that the Editorial Office might ask you for further information. Please add “The study was conducted according to the guidelines of the Declaration of Helsinki, and approved by the Institutional Review Board (or Ethics Committee) of NAME OF INSTITUTE (protocol code XXX and date of approval).” OR “Ethical review and approval were waived for this study, due to REASON (please provide a detailed justification).” OR “Not applicable” for studies not involving humans or animals. You might also choose to ex-clude this statement if the study did not involve humans or animals.

\informedconsent{Not applicable}%Any research article describing a study involving humans should contain this statement. Please add “Informed consent was obtained from all subjects involved in the study.” OR “Patient con-sent was waived due to REASON (please provide a detailed justification).” OR “Not applicable” for studies not involving humans. You might also choose to exclude this statement if the study did not involve humans. Written informed consent for publication must be obtained from participating patients who can be identified (including by the patients themselves). Please state “Written informed consent has been obtained from the patient(s) to publish this paper” if applicable.

\acknowledgments{\textls[-15]{We thank Zheng-Xiang Li, Yang-Jie Yan, Ji-Ping Dai, and Ran Gao for useful \mbox{discussions}.}}

\conflictsofinterest{The authors declare no conflicts of interest.} 

\begin{adjustwidth}{-\extralength}{0cm}
\printendnotes[custom] % Un-comment to print a list of endnotes

\reftitle{References}

%%%%%%%%%%%%%%%%%%%%%%%%%%%%%%%%%%%%%%%%%%
\end{adjustwidth}
\end{document}